\documentclass{emulateapj}
\usepackage{natbib}
\shorttitle{Accretion of the Magellanic System}
\shortauthors{Nichols et~al.}

\newcommand{\HI}{\ifmmode{{\rm H\scriptstyle I}}\else{H${\scriptstyle\rm I}$}\fi}

\def\gta{\;\lower 0.5ex\hbox{$\buildrel > \over \sim\ $}}
\def\lta{\;\lower 0.5ex\hbox{$\buildrel < \over \sim\ $}}

\begin{document}

\title{Accretion of the Magellanic system onto the Galaxy}
\author{Matthew Nichols}
\email{m.nichols@physics.usyd.edu.au}
\affil{Sydney Institute for Astronomy, School of Physics, The University of Sydney, NSW 2006, Australia}
\author{James Colless}
\affil{Sydney Institute for Astronomy, School of Physics, The University of Sydney, NSW 2006, Australia}
\author{Matthew Colless}
\affil{The Australian Astronomical Observatory, Epping NSW 1710, Australia}
\and
\author{Joss Bland-Hawthorn}
\affil{Sydney Institute for Astronomy, School of Physics, The University of Sydney, NSW 2006, Australia}

\begin{abstract}

Our Galaxy is surrounded by a large family of dwarf galaxies of which the most massive are the Large and Small Magellanic Clouds (LMC and SMC).
Recent evidence suggests that systems with the mass of the Local Group accrete galaxies in smaller groups rather than individually.
If so, at least some of the Galaxy's dwarfs may have fallen in with the LMC and SMC, and were formed as part of the Magellanic system in the nearby universe.
We use the latest measurements of the proper motions of the LMC and SMC and a multicomponent model of the Galactic potential to explore the evolution of these galaxy configurations under the assumption that the Magellanic system may once have contained a number of bound dwarf galaxies.
We compare our results to the available kinematic data for the local dwarf galaxies, and examine whether this model can account for recently discovered stellar streams and the planar distribution of Milky Way satellites.
We find that in situations where the LMC and SMC are bound to the Milky Way, the kinematics of Draco, Sculptor, Sextans, Ursa Minor and the Sagittarius Stream are consistent with having fallen in along with the Magellanic system.
These dwarfs, if so associated, will likely have been close to the tidal radius of the LMC originally and are unlikely to have affected each other throughout the orbit.
However there are clear cases, such as Carina and Leo~I, that cannot be explained this way.
\end{abstract}

\section{Introduction}
Most of the mass in our Universe is believed to be an unknown, collisionless form of matter that can only interact gravitationally with baryonic matter and itself \citep{Efstathiou1990}.
The formation of large-scale structures in our Universe, in particular galaxy clustering, is driven by gravitational forces exerted by this dark matter.
Cold Dark Matter (CDM) cosmology holds that structures grow hierarchically, with small objects collapsing first and then merging to form more massive galaxies and clusters.
The theory predicts that only dark matter halos with mass smaller than approximately $10^8$~M$_\odot$ can form from 3$\sigma$ fluctuations \citep{Read2006} in primordial density perturbations.
Consequently, more massive systems can only form by subsequent accretion of these protogalactic fragments.
For this reason, dwarf galaxies (dark matter subhalos in CDM cosmology) that contain luminous baryons and have not yet merged with a host galaxy, can, to some extent, be considered some of the most primitive building blocks of our universe.
However, it is now thought \citep{Tosi2003} that the inconsistencies between the observed properties of large galaxies and dwarfs are too many to believe that the former are built up only by means of successive accretions of the latter.
We investigate two key issues identified with this model.

\begin{figure}
  \centering
  \includegraphics[width=0.5\textwidth]{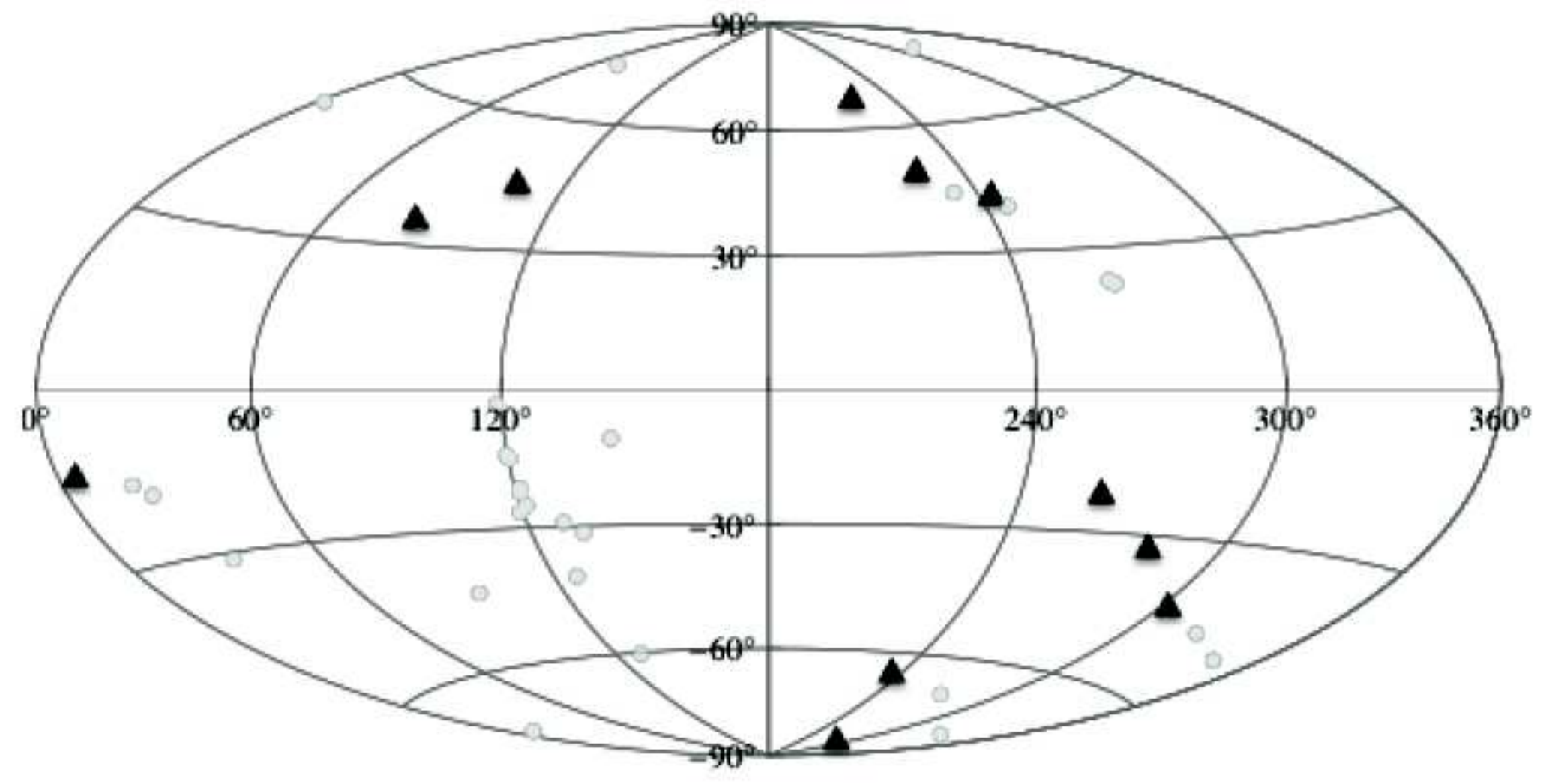}
  \caption{Hammer-Aitoff equal-area projection of the dwarf galaxy positions, with the 11 innermost Milky Way satellites highlighted (black triangles).}\label{fig:HAproj}
\end{figure}

The first issue has long been known, namely that the spatial distribution of the Milky Way satellites shows asymmetric patterns and probably streams \citep{Kunkel1976,Lynden-Bell1976,Hartwick2000}, particularly when only the innermost satellites are taken into account (Figure \ref{fig:HAproj}).
The orbits of these satellites have been found to be preferentially polar \citep{Zaritsky1999} based on wealth of evidence including the alignment of satellites on the sky \citep{Kunkel1976,Lin1982}, the orientation of the Magellanic Stream \citep{Mathewson1974}, the three-dimensional distribution of satellites \citep{Trevese1994} and their actual velocities \citep{Odenkirchen1994}.
\citet{Knebe2004} similarly find that nearby clusters such as Virgo and Coma possess galaxy distributions that tend to be aligned with the principal axis of the cluster itself.
They conclude that either some dynamical process is responsible or that the orbital parameters of the dwarf galaxies are imprinted on them at the time they enter the host halo.
They concluded that this hypothesis can be excluded at a 99.5\% level, given the empirical constraints showing the Milky Way potential to be spherical \citep{Fellhauer2006}.
Multiple arguments have been put forward to explain the disk-of-satellites problem, primarily based on CDM simulation models.
\citet{Kang2005} argue that the observed distribution of the Milky Way satellites is indeed consistent with being CDM subhalos, assuming that the satellites follow the distribution of the dark matter within the Milky Way halo.
\citet{Zentner2005} used a semi-analytic model to identify luminous satellites and showed that an isotropic distribution is not the correct null-hypothesis; rather, the host halos are mildly triaxial (tending to be more prolate than oblate).
Following similar lines to \cite{Libeskind2005}, they found that the distribution of the Galactic satellite system is in fact consistent with being CDM substructure, albeit with a low probability.
However, this disk-of-satellites is much less extended than nearby dwarf galaxy associations, and such an association falling into the Galaxy is unlikely to have produced the disk-of-satellites \citep{Metz2009}.

The second more specific problem is that, if simple accretion were the only mechanism for Galactic growth, then the spheroidal dwarf galaxy Leo~I has an improbably high radial velocity.
\citet{Mateo1998} finds the systemic velocity of Leo~I to be $287.0\pm1.9$~km~s$^{-1}$ for any subsample of the set of 33 red giants in the Leo~I dwarf spheroidal galaxy for which they obtained radial velocities from spectra taken using the HIRES echelle spectrograph on the Keck telescope.
This suggests that Milky Way satellites as a whole need to be considered in the wider context of the whole Local Group.
Due to this exceptional velocity, Leo~I has an anomalously large effect compared to other satellites on estimates of the mass of the Milky Way.
\citet{Sakamoto2002} use Bayes' theorem and the assumption of isotropic orbits for the satellites of our Galaxy to estimate a median probable mass of $\sim$$2.5\times10^{12}$~M$_\odot$.
Interestingly, however, the exclusion of Leo~I would lower this estimate to $\sim$$1.8\times10^{12}$~M$_\odot$.
Their result implies that Leo~I may not be gravitationally bound to the Milky Way (in order to agree with other estimates of the Milky Way mass derived by \citet{Wilkinson1999} and others, see Table~\ref{Tab:LMCSMCorb}).

In this paper we postulate an additional mechanism to explain the asymmetric dwarf galaxy distribution and the unusually high radial velocity of Leo~I.
We conjecture that some of the dwarf galaxies may have `piggybacked' in with the LMC-SMC binary pair, but had their bound orbits disrupted by interactions with the central Galactic potential.
Dwarfs bound to the LMC initially will be closer than most dwarf galaxy associations, and the subsequent interactions may have resulted in a 3-body interaction that ejected Leo~I and thus could reproduce its current high velocity while still explaining the narrow spacing of the disk-of-satellites association.
To test this hypothesis we first explore the past orbits of the LMC-SMC under the effect of a central Galactic potential pair in accordance with the latest measurements of their proper motions.
Numerically solving Newton's equations backwards in time, we proceed to find their position in phase space at apogalacticon. We then model the infall of this configuration as an {\em N}-body system containing the LMC and SMC as well as multiple dwarf galaxies (hereafter, all references to dwarfs and dwarf-dwarf interactions exclude the LMC and SMC) in bound orbits around the LMC.
We evolve our system forwards in time and see whether it can reproduce the anomalies in the Local Group discussed above, and if so what was the state of the dwarfs when bound to the Magellanic system.

\begin{deluxetable*}{lccc}
\tablecolumns{4}
\tablecaption{Orbital parameters for the LMC and SMC adopted or derived from
  sources as referenced.\label{Tab:LMCSMCorb}}
\tablehead{\colhead{}&\colhead{LMC}&\colhead{SMC}&\colhead{References}}
\startdata
Line of sight velocity (km~s$^{-1}$)  & $262.1\pm3.4$ & $146\pm0.6$ & \citet{Harris2006}\\
Proper Motions (W,N) (mas~yr$^{-1}$) & $-2.03 \pm 0.08, 0.44 \pm 0.05$ &$-1.16 \pm 0.18, -1.17 \pm 0.18$ & \citet{Kallivayalil2006}\\
Current Positions ($\alpha,\ \delta$) (deg)  & $81.9\pm0.3,-69.9\pm0.3$ & $13.2\pm0.3, -72.5\pm0.3$ & \citet{vanderMarel2002}\\
Galactic Coordinates (l, b) & $280.5, -32.5$ & $302.8, -44.6$ & - \\
Current positions (X, Y, Z) (kpc) & $-0.8, -41.5, -26.9$ & $15.3, -36.9, -43.3$ & - \\ 
Velocities (v$_{\rm X}$ , v$_{\rm Y}$ , v$_{\rm Z}$) (km~s$^{-1}$) & $-86 \pm 12$, $-268 \pm 11$, $252 \pm 16$ & $-87 \pm 48$, $-247 \pm 42$, $149 \pm 37$ & - \\
GC radial velocities (km~s$^{-1}$) & $89 \pm 4$ & $23 \pm 7$ & - \\
GC tangential velocities (km~s$^{-1}$) & $367 \pm 18$ & $301 \pm 52$ & - \\
\enddata
\end{deluxetable*}

\section{Fiducial Model}

The orbits of the LMC and SMC will have evolved over time as they interact gravitationally between each other and with the Galaxy over many gigayears.
Assuming minimal perturbation of the orbits of the LMC and SMC; the orbits of dwarfs that accompanied the Magellanic system can be calculated from a previous apogalacticon to the present day and into the future.

This is achieved by running a Monte Carlo suite of these multi-system models with varying initial conditions of the LMC and SMC (based upon the observed initial parameters, see \S\ref{ssec:LMCorb}).
The LMC/SMC orbits are calculated within a fixed Galactic potential and with gravitational interactions between the LMC and SMC (\S\ref{ssec:Pot}) backwards in time for $3.5$~Gyr and calculating the position of either the last or second last apogalacticon of the LMC.
At this apogalacticon a number of dwarfs are inserted in a bound orbit of the LMC (see \S\ref{ssec:DwarfIC}) and then traced forward in time to the present day and to $+0.5$~Gyr in the future, interacting with other dwarfs, the LMC/SMC system and the Galaxy.

\subsection{LMC and SMC orbital parameters}\label{ssec:LMCorb}
The orbit of the LMC and SMC will be affected by their current positions, velocities and by their masses.
The position and velocities used here for the LMC and SMC are given in Table~\ref{Tab:LMCSMCorb}.
Particular note should be taken of most recent measurements of the proper motions of the LMC and SMC by \citet{Kallivayalil2006}, which give a relative velocity between the clouds at the current epoch of $105\pm42$~km~s$^{-1}$. These values imply that the LMC tangential velocity is approximately $100$~km~s$^{-1}$ higher than previously thought \citep{Kallivayalil2006}.

Given the low uncertainty in position, we assume that the LMC and SMC are at fixed locations given by the best estimates of observations.
The velocity, with higher uncertainty, is randomly altered for each run assuming independent Gaussian uncertainties in the measurements, the line of sight velocity and the proper motions are calculated and then transformed to Cartesian velocities around the Galaxy \citep[for transformations to Cartesian coordinates see][]{vanderMarel2002}

The masses of the LMC and SMC have larger uncertainties than that of the position and velocity.
\citet{vanderMarel2002} obtained a mass for the LMC of $(8.7\pm{}4.3)\times10^{9}$~M$_\odot$ within $8.9$~kpc using an analysis of carbon stars.
This mass is less than half that estimated by \citet{Schommer1992} who derive a mass of $2\times10^{10}$~M$_\odot$ using radial velocities for several of the oldest star clusters in the LMC that lie well beyond $6$~kpc of its center.
We match the circular velocities of \citet{vanderMarel2002} with that of an Einasto halo at $8.9$~kpc to get a virial mass of $2\times10^{11}$~M$_\odot$ for a halo virialized today.
Using a simplified model of tidal radius $r_{\rm tid}^3/R_{\rm peri}^3 = m_{\rm tid}/M_{\rm Galaxy}$ we find that the LMC will have a mass of $5.6\times10^{10}$~M$_\odot$ within a tidal radius of $30$~kpc, only slightly more massive than that of \citet{Schommer1992}.

For the SMC, \citet{Hardy1989} obtained a lower mass limit of $1.0\times10^{9}$~M$_\odot$ from observations of carbon stars, while \citet{Dopita1985} obtained $0.9\times10^{9}$ from similar observations of planetary nebulae.
More recently, \citet{Harris2006} determined a mass of $2.7$--$5.1\times10^{9}$~M$_\odot$.
We use the velocity dispersion from \citet{Harris2006} to calculate an Einasto halo virial mass of $4.5\times10^{10}$~M$_\odot$, with a mass $5.7\times10^9$~M$_\odot$ within a tidal radius of $19$~kpc.

These masses, somewhat above the calculated masses at smaller radii, also approximate the mass of a theorized common halo between the LMC and SMC \citep[see for example][]{Bekki2008}, without the problem of a large subhalo within subhalo (in common halo models, the LMC halo will have a mass of $\sim$$20\%$ of the common halo).
These masses are also similar to those calculated from the stellar mass content and seen inside {\em N}-body simulations \citep{Boylan-Kolchin2011}.

For the model, we ignore any mass loss from the clouds (in particular that due to tidal stripping), although mass loss is known to occur as evidenced by the amount of matter in the Magellanic Stream; \citep[$\sim$$2\times10^8$~M$_\odot$][]{Putman2003}.
Dynamical friction between the LMC, SMC and the Galaxy is also calculated using the tidal masses of the LMC and SMC.
The effect of dynamical friction between the LMC and SMC is ignored, although LMC/SMC binary systems are still found in {\em N}-body simulations that include dynamical friction \citep{Boylan-Kolchin2011} suggesting that the survival time with dynamical friction is sufficiently large for the Magellanic system to reach the present day with these masses.

\subsection{Potentials}\label{ssec:Pot}
The LMC/SMC and dwarf galaxy orbits occur within the potential of the Galaxy and are altered due to interactions between the Magellanic clouds themselves and the dwarfs.

There has been much work on Galaxy potential models since the first basic schemes devised by \citet{Murai1980} and \citet{Fujimoto1976}.
In contrast to other recent approaches that take simple spherical Galactic potentials, we choose to model the Galaxy using a multicomponent model of the Galactic potential developed by \citet{Flynn1996}.
This model matches known Galactic parameters such as the rotation curve, local disk density and disk scale-length to high accuracy.
The lack of noticeable warp in the Magellanic Stream either in position on the sky or radial velocity---see \citet{Lin1995}---suggests that the disk and bulge likely has small or negligible effects on the orbits of the clouds, allowing us to have greater faith in our derived orbits.
In this model the Galactic potential $\Phi(R,z)$ is given in cylindrical coordinates, where $R$ is the planar Galactocentric radius, and $z$ is the distance above the plane of the disk. The total potential $\Phi$ is modelled by the sum of the three different potentials: the dark halo $\Phi_H$, a central component $\Phi_C$, and a disk $\Phi_D$; thus
\begin{equation}
\Phi=\Phi_H+\Phi_C+\Phi_D.
\end{equation}

The potential of the dark halo $\Phi_H$ is assumed to be spherical and of the form 
\begin{equation}
  \Phi_H=\frac{1}{2}V_H^2\ln(r^2+r_0^2)
\end{equation}
where $r$ is the Galactocentric radius $(r^2=R^2+z^2)$.
The potential has a core radius $r_0$, and $V_h$ is the circular velocity at large $r$.
The potential of the central component $\Phi_C$ is modelled by two spherical components, representing the bulge/stellar-halo and inner core components:
\begin{equation}
\Phi_C=\frac{GM_{c1}}{\sqrt{r^2+r_{c1}^2}}-\frac{GM_{c2}}{\sqrt{r^2+r_{c2}^2}}
\end{equation}
where $G$ is the gravitational constant, $M_{c1}$ and $r_{c1}$ are the mass and core radius of the bulge/stellar-halo term, and $M_{c2}$ and $r_{c1}$ are the mass and core radius of the inner core.
The disk potential $\Phi_D$ is modelled using an analytical form that is a combination of three Miyamoto-Nagai potentials \cite{Miyamoto1975}:
\begin{equation}
\Phi_D=\Phi_{D1}+\Phi_{D2}+\Phi_{D3}
\end{equation}
where
\begin{equation}
\Phi_D=\frac{-GM_{D_n}}{\sqrt{(R^2+[a_n+\sqrt{(z^2+b^2)}]^2)}} ~~~~ n=1,2,3...
\end{equation}
and the parameter $b$ is related to the disk scale-height, $a_n$ to the disk scale-length, and $M_{D_n}$ the masses of the three disk components.
\citet{Flynn1996} discuss in detail the justification of this model and its close fit to observational data.
We assume hereafter that it reasonably accurately models the Galactic potential with the adopted parameters listed in Table \ref{Tab:parameters}.

\begin{deluxetable}{lcc}
\tablecaption{Adopted parameters for the Galactic potential.\label{Tab:parameters}}
\tablecolumns{3}
\tablewidth{0pt}
\tablehead{\colhead{Component}&\colhead{Parameter}&\colhead{Value}}
\startdata
Dark Halo  & $r_0$ & $8.5$~kpc\\
   & $V_H$ & $220$~km~s$^{-1}$\\
Bulge/Stellar-halo &  $r_{C_1}$ & $2.7$~kpc\\
   & $M_{C_1}$ & $3.0 \times 10^9$~M$_\odot$\\
Central Component &  $r_{C_2}$ & $0.42$~kpc\\
   & $M_{C_2}$ & $1.6 \times 10^{10}$~M$_\odot$\\
Disk &  $b$ & $0.3$~kpc\\
 & $M_{D_1}$ & $6.6 \times 10^{10}$~M$_\odot$\\
&  $a_1$ & $5.81$~kpc\\
 & $M_{D_2}$ & $-2.9 \times 10^{10}$~M$_\odot$\\
&  $a_2$ & $17.43$~kpc\\
 & $M_{D_3}$ & $3.3 \times 10^9$~M$_\odot$\\
&  $a_3$ & $34.86$~kpc\\
\enddata
\end{deluxetable}

Solving Poisson's equation with the Galactic potential and integrating to $100$~kpc gives a total mass of $M_{\rm{MW}}=1.43\times10^{12}$~M$_\odot$.
This value agrees well with Galactic mass derived in previous work for radii greater than 20~kpc (see Table \ref{Tab:mass-radius}).

\begin{deluxetable}{lcc}
\tablecaption{Comparison of masses between various Milky Way models at radii
  of 50~kpc and 100~kpc.\label{Tab:mass-radius}}
\tablecolumns{3}
\tablehead{\colhead{Model}&\colhead{$M_{50} (10^{11}$~M$_\odot)$} & \colhead{$M_{100} (10^{12}$~M$_\odot)$}}
\tablewidth{0pt} 
\startdata
\textbf{Adopted}  & \textbf{4.3} & \textbf{1.4}\\
\citet{Wilkinson1999}        &  5.4 & 1.9\\
\citet{Kochanek1996}  & 4.9 & ...\\
\citet{Li2008} & ...  & 1.81\\
\citet{Smith2007}  & ... & 0.7-2.0\\
\citet{Sakamoto2002}  & 5.4 & 1.8\\
\enddata
\end{deluxetable}

Although the potential well of the Galaxy is by far the deepest, the LMC, SMC and dwarfs will interact with each other.
We assume that the LMC and SMC (and each dwarf) consist of an Einasto halo, with a sharp boundary at the tidal radius.
The acceleration experienced by one of these objects at ${\bf x}_{1}$ by another at ${\bf x}_{2}$, separated by a distance $r=||{\bf x}_1 - {\bf x}_2||$ is then \citep{Nichols2009}
\begin{equation}
  \frac{\rm{d}\Phi}{\rm{d}{\bf x}} =\left\{ 
\begin{array}{ll}
3v_s^{2}r_s2^{-2-3/\alpha}\exp(2/\alpha)\alpha^{-1+3/\alpha}\\
\times\quad\gamma(3/\alpha,2[r/r_s]^{\alpha}/\alpha)\frac{({\bf x}_1 - {\bf x}_2)}{r^3} &{\rm if~} r < r_{{\rm tidal},2}\\
\\
\frac{GM_{2}({\bf x}_1 - {\bf x}_2)}{r^{3/2}} &{\rm if~}  r \ge{}r_{{\rm tidal},2},
\end{array} \right.
\end{equation}
where $v_s$ is the halo scale velocity, $r_s$ the halo scale radius and $\alpha$ the Einasto scale factor (set here at $\alpha=0.18$).

\subsection{Dwarf Galaxy Orbital Parameters}\label{ssec:DwarfIC}

The relatively narrow spread in dwarf galaxy positions suggests that if dwarf galaxies did fall in with the Magellanic system they would be more tightly bound than most dwarf associations seen today.
We hence assume that the dwarfs were in a bound orbit of the most massive object in the Magellanic system, the LMC, at a previous apogalacticon of the LMC.

Each dwarf is randomly assigned a periapsis and circularity (and hence eccentricity) according to the distributions of \citet{Wetzel2011} extended down to the lower mass of the LMC.
These distributions lead to a CDF for periapsis of
\begin{eqnarray}
  F\left(\frac{r_{\rm peri}}{{\rm kpc}}\right) &=&  48.7\{\gamma^{-1}(3/17,\gamma_{\rm min}\nonumber\\
  &+&[r_{\rm peri}/r_{\rm vir, LMC}]^{0.85}[\gamma_{\rm max} - \gamma_{\rm min}])\}^{1/0.85},
\end{eqnarray}
where, $\gamma^{-1}$ is the inverse lower incomplete gamma function and 
\begin{equation}
  \gamma_{\rm min/max} = \gamma(3/17,[0.32r_{\rm min/max}/r_{\rm vir, LMC}]^{0.85}),
\end{equation}
where $r_{\rm min/max}$ is the minimum and maximum allowed periapsis of the dwarfs.
We choose a $r_{\rm max}$ of the tidal radius, as any dwarf that finds itself outside the tidal radius at closest approach will experience a greater attraction to the Milky Way than the LMC and will hence be definitely lost.
A minimum periapsis is chosen to be three times the exponential scale length of the LMC disk \citep[disk scale length $\sim1.4$~kpc][]{vanderMarel2002}, orbits close to the disk are likely to produce warping that is not observed within the LMC disk.

The circularity CDF is given by
\begin{equation}
  F(\eta) = 2.25\eta^{2.05} \phantom{.}_2F_1(2.05,-0.66;3.05;\eta),
\end{equation}
where $\eta$ is the circularity of the orbit and $\phantom{.}_2F_1(a,b;c;z)$ is the Gaussian hypogeometric function.

Any orbits which would extend past the calculated tidal radius are excluded and a new one calculated.

The position of the dwarf within the orbit is determined by approximating the rosette orbit that arises from the spherical potential as a Keplerian orbit \citep[the distributions also approximate it as Keplerian, but an extended NFW halo sees no significant change outside of low periapsis orbits, see][]{Wetzel2011} and is found by solving
\begin{eqnarray}
P\left(r < xr_{\rm peri}\right) &=& \frac{1}{\pi}\left(\frac{\pi}{2} - \beta - \arctan\left[\frac{1-x(1-\epsilon)}{\beta}\right]\right),\\
  \beta &=& \sqrt{(x-1)(1-\epsilon)(1+\epsilon-x[1-\epsilon])}\nonumber, 
\end{eqnarray}
where $\epsilon=\sqrt{1-\eta^2}$ is the eccentricity of the orbit.

By randomly determining the angular position of the orbit and the radial direction of travel, the position and velocity can then be determined in Cartesian coordinates.

The mass of the dwarfs was set at either $1.1\times10^7$~M$_\odot$ within a tidal radius of $300$~pc, consistent with the common mass of dwarf halos from \citet{Strigari2008} or a mass of $2.36\times10^8$~M$_\odot$ within the same tidal radius which represents a dwarf halo ten times more massive at the virial radius.
The tidal radius of the dwarfs was set arbitrarily to enclose this mass, although any dependence on mass is small (\S\ref{ssec:dwnum}).

\section{Model parameters and their effects}
In addition to the randomized position and velocity of the LMC/SMC and the dwarfs; the number of dwarfs bound to the LMC, the mass of these dwarfs, and when the dwarfs were bound to the LMC were altered.
These parameters were altered throughout five different models, with the parameters shown in Table \ref{Tab:models}.
Model F, which had no apogalacticon was calculated by multiplying the calculated error (see \S\ref{ssec:LMCorb}) by a factor of three, in order to achieve the required velocities to be unbound to the Galaxy under this potential.

\begin{deluxetable}{lcccccc}
\tablewidth{0pc}
\tablecolumns{7}
\tablecaption{Group Infall Models. \label{Tab:models}}
\tablehead{\colhead{} & \colhead{} & \colhead{M$_{\rm dwarf}$} & \colhead{Initial} & \colhead{Number} & \multicolumn{2}{c}{\% Unbound}\\
\colhead{Model} & \colhead{$N_{\rm D}$} & \colhead{($10^{7}$~M$_\odot$)}& \colhead{Apo} & \colhead{of runs} & \colhead{Today} & \colhead{$+500$~Myr}}
\startdata
A & $1$ & $1.1$ & $1$ & $27800$ & $28\%$ & $37\%$\\
B & $1$ & $1.1$ & $2$ & $5600$ & $87\%$ & $87\%$\\
C & $4$ & $1.1$ & $1$ & $25200$ & $28\%$ & $37\%$\\
D & $4$ & $1.1$ & $2$ & $26300$ & $87\%$ & $87\%$\\
E & $4$ & $23.6$ & $1$ & $3900$ & $26\%$ & $36\%$\\
F & $4$ & $1.1$ & $\star$ & $2500$ & $8\%$ & $9\%$
\enddata
\tablecomments{$N_{\rm D}$ is the number of dwarfs originally bound to the LMC, Initial Apo is the number of apogalacticons before the present from which the orbits are tracked to the present day and the percentage unbound is the fraction of dwarfs that have enough energy to escape from an isolated LMC.
Model F had no apogalacticon and was begun $2500$~Myr ago.}
\end{deluxetable}

\subsection{The effects of the mass and number of dwarfs}\label{ssec:dwnum}
The amount of dwarfs will determine the amount of dwarf-dwarf interactions that occur, and within the comparatively small halo of the LMC these dwarfs could gravitationally disturb each other.
The frequency of interactions between small dwarfs (with a mass $M_{\rm dwarf}/M_{\rm LMC}\sim10^{-3}$) bound to the LMC will be a factor of their orbital distribution---if they themselves are in a disk-of-satellites around the LMC, not modelled here, interactions will be more frequent---and the amount of dwarfs present.

The frequency and effect of these dwarf-dwarf interactions are examined by comparing the final energies of the dwarfs around the LMC in models A and C and models B and D.
We consider a dwarf to be unbound at a point in time if it has enough energy such that it would escape the LMC considering only a two-body interaction.
A two-sample Kolmogorov--Smirnov test could not rule out the null hypothesis that the distributions in final energies from the LMC of models A and C or models B and D are consistent today (with probabilities that the distributions are consistent of $p_{\rm AC}>0.3$, $p_{\rm BD}>0.4$) or at $+500$~Myr ($p_{\rm AC}>0.3$, $p_{\rm BD}>0.9$).
This null result suggests that dwarf-dwarf interactions are rare with only a few dwarfs bound to the LMC, or that such interactions cause only slight perturbations in other bound dwarfs orbits.

The effect of increasing dwarf tidal mass by a factor of $\sim$$20$ is measured by the comparison of the energy of the dwarfs in models C and E.
Applying a two-sample Kolmogorov-Smirnov test to the energy of the dwarfs from models C and E suggests the null hypothesis cannot be ruled out with probabilities of the same distribution being $p_{\rm CE}\sim0.03$ today and $p_{\rm CE}>0.2$ at $+500$~Myr.
The increase in $p$ value at $+500$~Myr for all model comparisons may be a consequence of the strongest LMC-Galaxy-dwarf interactions occurring near perigalacticon, approximately today, with any variance along the LMC-SMC orbit having a bigger effect now than in $+500$~Myr.

\subsection{Initial time effects}
Due to the large differences in the percentage of dwarf galaxies that become unbound between models that began at the last or second last apogalacticon or that had no apogalacticon, see Table \ref{Tab:models}, it would be expected that there are significant variations in the energy distributions of the dwarfs.
Using a two sample Kolmogorov-Smirnov test shows significant differences between the final energy of dwarfs in models A and B today with the probability of the null hypothesis, that is the the final energies of dwarfs in models A and B coming from the same distribution, being $p_{\rm AB} < 10^{-3}$ with a KS statistic of $D = 0.70$ and between models C and D today $p_{\rm CD} < 10^{-3}$ with a KS statistic of $D = 0.69$.
These energy distributions are also significant between models C and F ($p_{\rm CF}<10^{-3}$, $D=0.20$) and models D and F ($p_{\rm DF} < 10^{-3}$, $D=0.81$).  
Similarly the differences are significant at $+500$~Myr ($p < 10^{-3}$ between  A\&B, C\&D, C\&F and D\&F).
Even within bound dwarfs, the distributions in energy are significantly different ($p < 10^{-3}$ A \& B, C \& D, C \& F and D \& F).
As a run lost a higher percentage of dwarfs the remaining bound dwarfs were of lower energy; that is, bound dwarfs in model F were of higher energy than dwarfs in model C which were of higher energy than model D.
This loss of high energy dwarfs, occurs as the LMC/SMC approaches perigalacticon.
Runs in model F, which had orbits most altered by dwarfs due to the large timescales, were more likely to lose dwarfs the closer the perigalacticon passage (occurring approxiamtely today).

\section{Results}

As the ranges of mass and number of dwarfs explored show little effect on the number of ejected systems, we focus on the orbits of the LMC/SMC and the orbits of the dwarfs in models C, D and F in order to examine these effects.

\subsection{Orbits of the LMC and SMC}\label{ssec:LMCSMCorbs}

\begin{figure*}
\includegraphics[width=\textwidth]{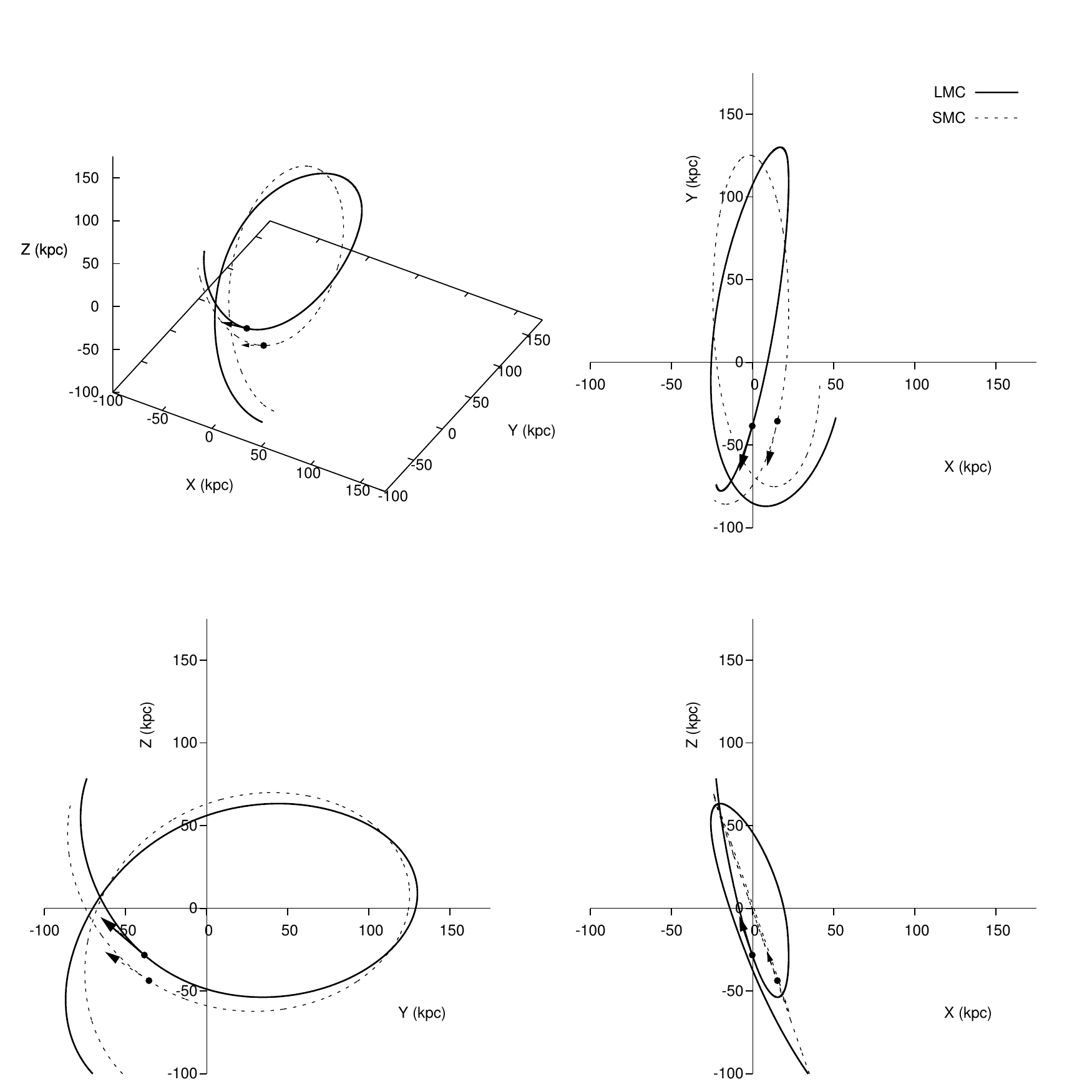}
\caption{An orbit where the SMC is bound to the LMC from model D with current position as circles and velocities shown as vectors (size of vector in kpc is the velocity in $10$'s of km~s$^{-1}$). The orbit of the LMC is shown as solid line and the orbit of the SMC is shown as a dashed line.\label{fig:orb}}
\end{figure*}

\begin{figure*}
\includegraphics[width=\textwidth]{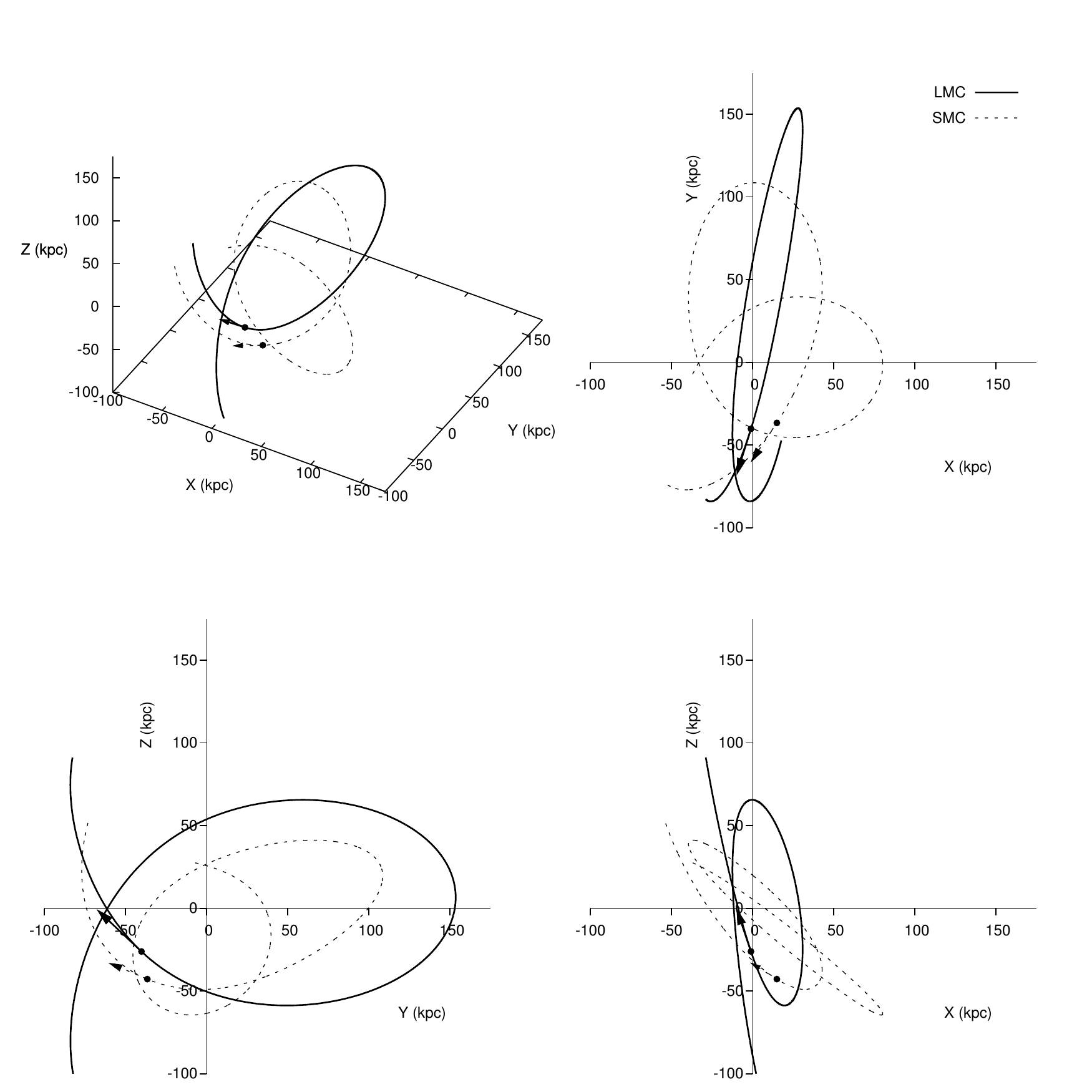}
\caption{An orbit where the SMC is not bound to the LMC from model D with current position as circles and velocities shown as vectors (size of vector in kpc is the velocity in $10$'s of km~s$^{-1}$). The orbit of the LMC is shown as solid line and the orbit of the SMC is shown as a dashed line.\label{fig:orb2}}
\end{figure*}

Due to the varying initial positions and velocities of the LMC and SMC, the orbits differ substantially between runs, with times for apogalacticons varying by hundreds of millions of years.
Within model C, the SMC is bound to the LMC at the last apogalacticon in $13840$ ($55\%$) runs, within model D the SMC is bound to the LMC two apogalacticons ago in $2691$ ($10\%$) runs and within model F the SMC is bound to the LMC $2500$~Myr ago in $559$ ($22\%$) runs.
A bound orbit for model D is shown in Figure \ref{fig:orb} from $t\sim-2500$~Myr, with the current velocity for both the LMC and SMC shown as a vector emanating from their current position.
An unbound orbit for model D is shown in Figure \ref{fig:orb2} from $t\sim-3000$~Myr, with the initial velocity again as a vector.
Due to the large number of orbits in which the SMC may be unbound, this orbit is not necessarily typical of all runs.

If the SMC was unbound at the last apogalacticon while the dwarfs were bound, the dwarfs were more likely to be ejected ($p<10^{-3}$), possibly as a result of the SMC and LMC becoming bound.
However, this effect is small, with on average $0.08$ more dwarfs ejected per run in model C when the SMC was unbound at the last apogalacticon, with no impact if the SMC was bound at the second to last apogalacticon (model D, $p>0.05$).
This suggests that the SMC becoming bound to the LMC may only remove dwarfs that would in any case become unbound at a future point.

\subsection{The orbits of dwarfs}
Despite being gravitationally bound at a previous apogalacticon, a large fraction of dwarfs become unbound from the LMC, often ending up a great distance away from the Magellanic system at the present day.

\begin{figure*}
\includegraphics[width=\textwidth]{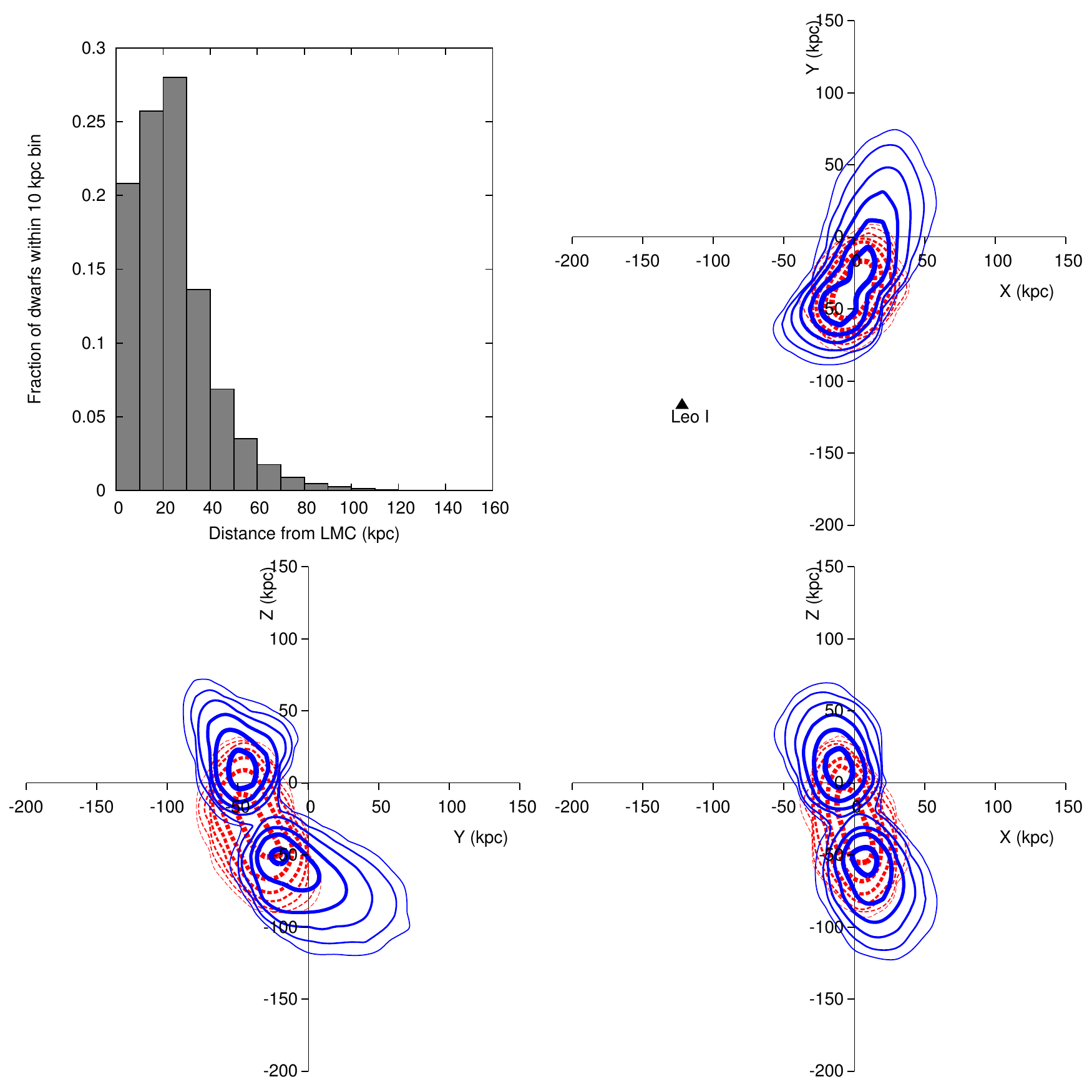}
\caption{The distance from the LMC of the dwarfs in model C at the present time and density contours of the final Cartesian location.
The contours of dwarfs that are unbound are shown as (blue) solid lines and dwarfs which remain bound as (red) dashed lines.
The contours give the fraction of all dwarfs within a $10$~kpc radius in seven logarithmic steps from $10^{-4}$ to $10^{-1}$ with the contours increasing in line width with increasing density.
Leo~I is plotted for reference, but is off the plot in Z (at $\sim$200~kpc).
\label{fig:fposC}}
\end{figure*}

\begin{figure*}
\includegraphics[width=\textwidth]{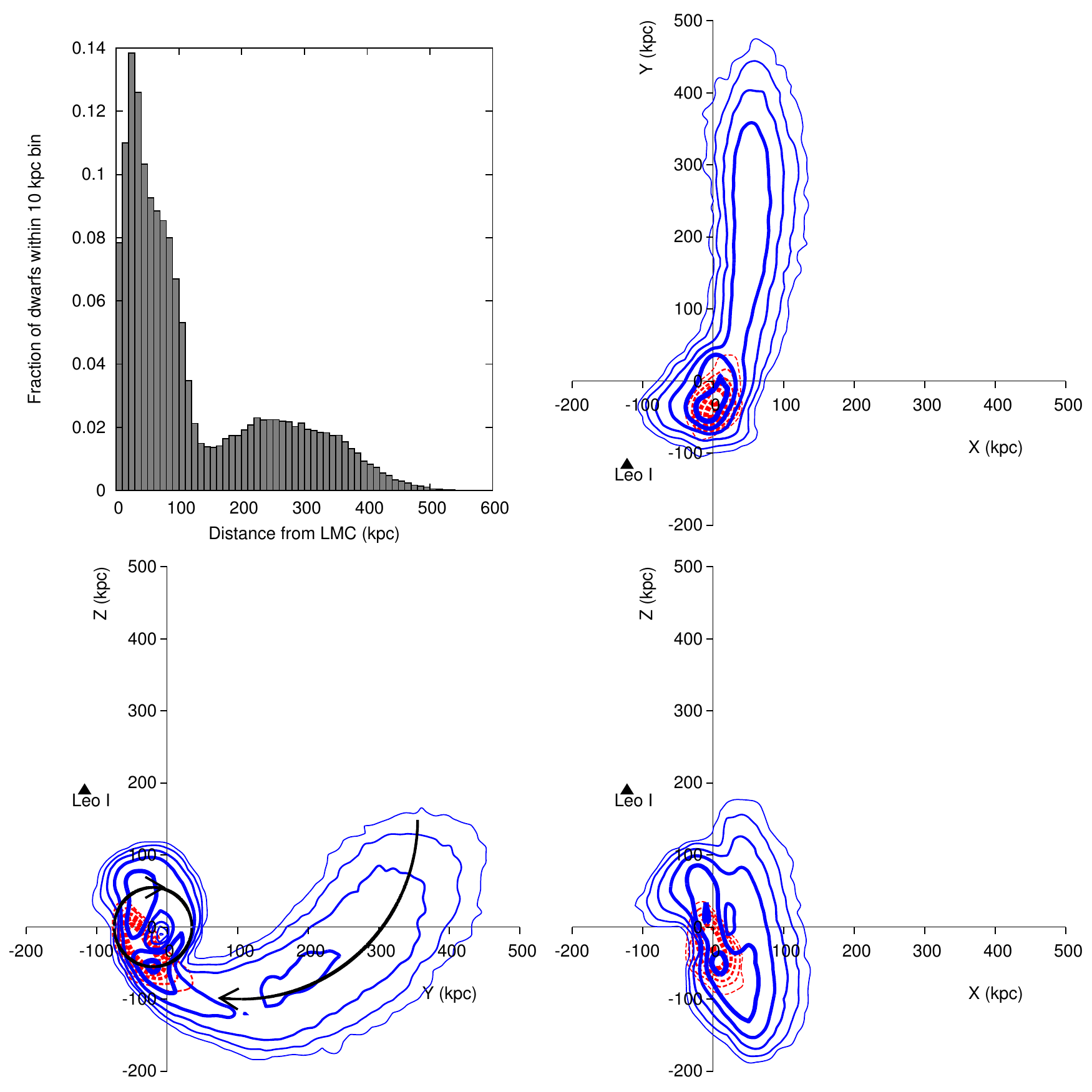}
\caption{The distance from the LMC of the dwarfs in model D at the present time and density contours of the final Cartesian location.
The contours of dwarfs that are unbound are shown as (blue) solid lines and dwarfs which remain bound as (red) dashed lines.
The contours show the fraction of all dwarfs within a $10$~kpc radius in seven logarithmic steps from $10^{-4}$ to $10^{-1}$ with the contours increasing in line width with increasing density.
The two unbound types of orbits for dwarfs are plotted schematically in the Y--Z plane.
Leo~I is plotted for reference.
\label{fig:fposD}}
\end{figure*}

\begin{figure*}
\includegraphics[width=\textwidth]{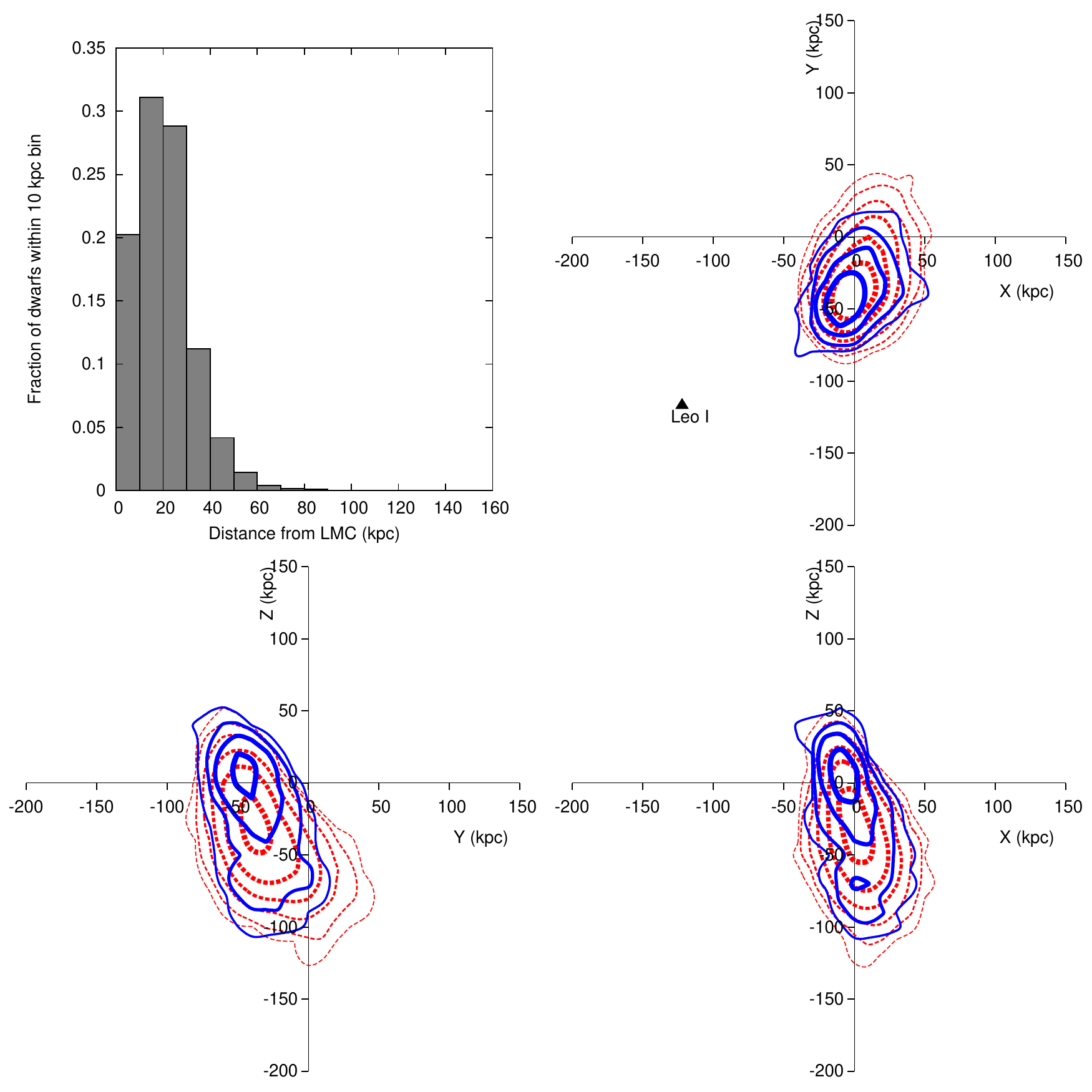}
\caption{The distance from the LMC of the dwarfs in model D at the present time and density contours of the final Cartesian location.
The contours of dwarfs that are unbound are shown as (blue) solid lines and dwarfs which remain bound as (red) dashed lines.
The contours show the fraction of all dwarfs within a $10$~kpc radius in six logarithmic steps from $10^{-3.5}$ to $10^{-1}$ with the contours increasing in line width with increasing density.
The two unbound types of orbits for dwarfs are plotted schematically in the Y--Z plane.
Leo~I is plotted for reference.
\label{fig:fposF}}
\end{figure*}

The final positions today (shown in Figure \ref{fig:fposC} for model C, Figure \ref{fig:fposD} for model D and Figure \ref{fig:fposF} for model F) show clear differences due to the starting apogalacticon.

For dwarfs that were bound at the last apogalacticon (model C), $\sim$$30\%$ of these will be unbound today, however, they remain relatively close to the LMC and are spread out along the orbital path of the Magellanic system.
These unbound dwarfs form a bimodal distribution leading and trailing the LMC.
This bimodality illustrates the difficulty for a dwarf to remain unbound near the LMC, whose gravity will recapture most dwarfs close by.

Of the dwarfs that were bound two apogalacticons ago (model D), a much larger fraction ($\sim$$90\%$) have become unbound by the present day and have begun their own orbits around the Galaxy.
These unbound dwarfs, occupy three main types of orbits, a relatively small ($\sim$$100$~kpc) low eccentricity orbit around the Galaxy, a much larger ($\sim$$300$~kpc) highly eccentric orbit, with others remaining close to the LMC/SMC and travelling along similar orbits.
As in model C, dwarfs that are close to the LMC and unbound have a bimodal distribution leading and trailing, again showing that is difficult for a dwarf to remain unbound near the LMCs gravitational influence.

Dwarfs that accompanied the LMC in an originally unbound orbit (model F), very few have become unbound in the model today or at $+500$~Myr.
These dwarfs are much more likely to have a significant effect on the orbit of the LMC and SMC and correspondingly show a larger region of bound dwarfs as the model LMCs orbit is altered from the initial backwards integration.
The distribution of unbound dwarfs is similar, however, to those dwarfs bound at the LMC at the last apogalacticon with a leading group and a slight overdensity trailing the LMC.

Despite these dwarfs being ejected from the system, they remain in the same plane as the orbit of the LMC and SMC and within the disk-of-satellites seen around the Galaxy today, these unbound dwarfs also end up retaining their original direction of travel despite potentially being ejected at high velocity in a complicated three-body interaction.
All but one dwarf from models C and D are moving clockwise in the Y--Z plane, with the one dwarf travelling anti-clockwise having only a small velocity in the plane ($v_{\rm YZ}\sim8$~km~s$^{-1}$).

Of note is that there are no dwarfs in models C, model D or F which comes close to the current position of Leo~I ($-122$~kpc,$-117$~kpc,$189$~kpc) \citep{Kroupa2005}.
In addition with Leo~I likely being unbound to the Galaxy, if it did infall bound to the Magellanic system an extremely rare multi-body reaction must have occurred to boost it away from the Galaxy.
Similar ejections have been seen in large {\em N}-body simulations \citep{Sales2007} between satellites of the host system.
Leo~I could therefore still have been associated with the Magellanic system, but unbound to the LMC when entering the Galactic halo.

The large fractions of simulated dwarfs that were bound in the past and unbound today means that the lack of dwarfs observed to be orbiting the LMC today does not imply there were no dwarfs bound to the Magellanic system in the past.

\subsection{Differences between bound and unbound dwarfs}

With a large fraction of dwarfs becoming unbound, many of which leave the Magellanic system, the properties that allow dwarfs to become unbound can be examined.
In particular, the contribution of the orbits of the LMC and SMC and the initial orbit of the dwarfs themselves around the LMC.

The effect of variations in the Magellanic system's orbit on dwarf loss is examined by looking at how many dwarfs were lost in each run.
If the LMC/SMC orbits have significantly different loss rates, then future refinement of the orbit of the clouds may give evidence against dwarfs falling in with the clouds.
If no physical property of the LMC/SMC orbits correlates significantly with dwarf ejection probabilities, then we can model ejections as a Poissonian distribution.
Hence if $\sim$30$\%$ of all dwarfs were lost, as in model C, and with no contribution from the orbits of the LMC and SMC, the number of runs which lose $0$ to $4$ dwarfs would be a Poissonian distribution with ratios of $(6748,10530,6162,1602,156)$.
In fact the runs from model C lose these dwarfs in the proportions of $(6866,10379,6137,1637,181)$, a difference that is statistically insignificant (a $\chi^2$-test gives $p>0.05$).
The effect of the orbits is also not seen after the second to last apogalacticon in model D; again modelling as a Poisson process the number of dwarfs lost per run would be expected to be in the proportions $(6,181,1895,8820,15397)$, while the results from model D give $(6,225,1891,8697,15481)$; (a $\chi^2$-test gives $p>0.01$).
This suggests that the positions of the dwarfs at apogalacticon are much more important than the orbits of the LMC and SMC (excluding whether the SMC is bound initially, see \S\ref{ssec:LMCSMCorbs}).

\begin{figure}
  \centering
  \includegraphics[width=0.5\textwidth]{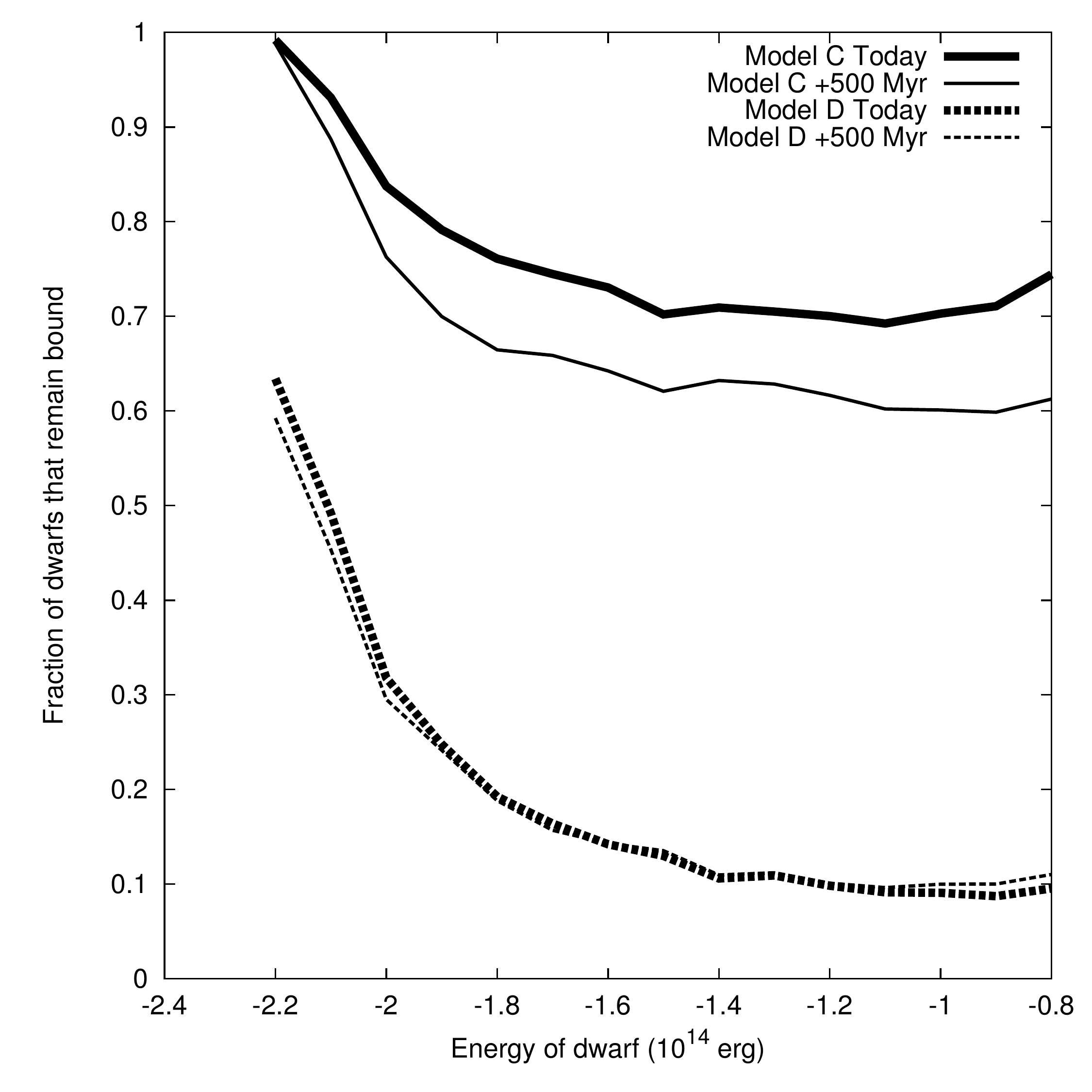}
  \caption{The fraction of dwarfs that remain bound as a function of their beginning energy.
A dwarf deep inside the potential well is unlikely to become unbound, as the orbit increases in energy up to $\sim$$-1.8\times10^{14}$~erg the dwarf is increasingly likely to become unbound.
Dwarfs that were bound two apogalacticons ago (model D) are just as likely to be stripped in $+500$~Myr as now, indicating that any stripping of the dwarfs from the Magellanic system will have already taken place.
Dwarfs that were bound one apogalacticon ago (model C) are still undergoing stripping from the Magellanic system, with dwarfs more likely to be unbound at $+500$~Myr then the present day.\label{fig:boundperc}}
\end{figure}

The impact of different orbits of dwarfs around the LMC at apogalacticon is also examined.
Dwarfs which originally possess a low energy---that is, are deep within the potential well of the LMC---are less likely to be stripped either when the SMC passes close by or at perigalacticon, when the gravitational attraction of the Galaxy is largest.
A Kolmogorov-Smirnov test comparing the initial energy of the dwarf orbits that remain bound and those that are unbound indicates that the distributions of initial energy are different for model C today and at $+500$~Myr (the null-hypothesis that the distributions are the same has probabilities of $p_{\rm today}<10^{-3}$ and $p_{+500~{\rm Myr}}<10^{-3}$) and for model D today and at $+500$~Myr ($p_{\rm today}<10^{-3}$, $p_{+500~{\rm Myr}}<10^{-3}$).
The fraction of dwarfs that remain bound as a function of initial energy is shown in Figure \ref{fig:boundperc}.
Dwarfs that were bound one apogalacticon ago (model C) are still undergoing removal from the Magellanic system as it passes close to the Galactic center.
Dwarfs that were bound two apogalacticons ago (model D) are no longer undergoing removal, with dwarfs that are bound now remaining bound over the next $500$~Myr, with a handful being recaptured as they pass close by the LMC during their own orbit of the Galaxy.
No matter when the dwarfs were originally bound, dwarfs that were deep inside the potential well originally are much more likely to be bound today and remain bound over the next $500$~Myr, while dwarfs of energy above about $-1.8\times10^{14}$~erg are all equally likely to be removed.

\begin{figure*}
  \includegraphics[width=\textwidth]{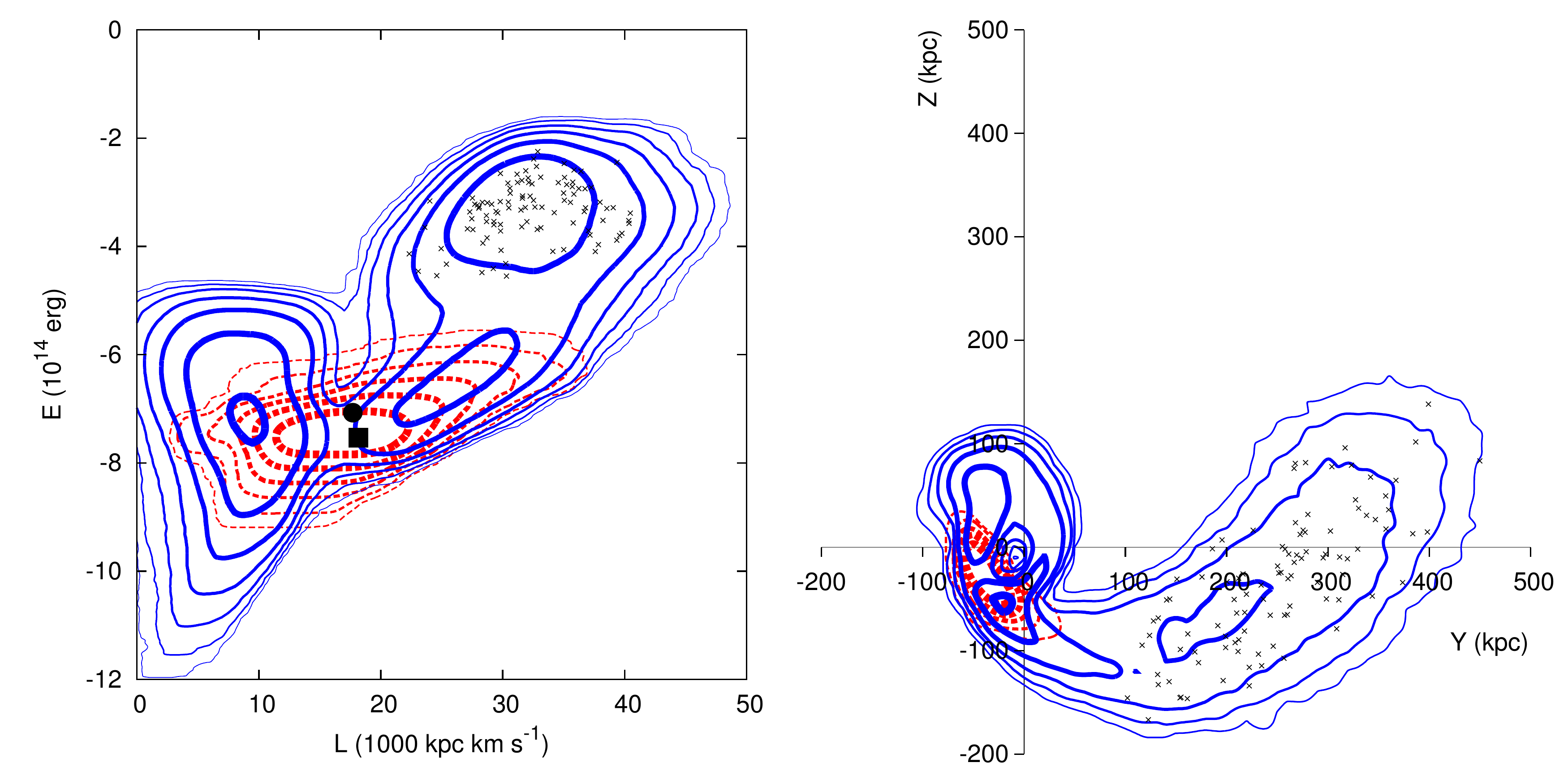}
  \caption{Energy-angular momentum Diagram of the dwarfs in model D.
Even though unbound and bound dwarfs are likely to be found in the same regions, they occupy different regions of the energy-angular momentum diagram.
Unbound dwarfs which have highly eccentric orbits around the Galaxy have higher energy for the same angular momentum than dwarfs on low eccentricity orbits.
The positions of the LMC and SMC on the energy-angular momentum diagram are plotted as a black square and circle respectively.
The contours in the energy-angular momentum diagram show the fraction of all dwarfs within a $5000$~kpc~km~s$^{-1}\times5\cdot10^{13}$~erg~g$^{-1}$ box, in seven logarithmic steps from $10^{-4}$ to $10^{-1}$ with the contours increasing in line width with increasing density.
The contours on the right show the fraction of all dwarfs within a $10$~kpc radius in seven logarithmic steps from $10^{-4}$ to $10^{-1}$ with the contours increasing in line width with increasing density.
The small crosses are a random selection of $100$ dwarfs from the large eccentricity tale (Y$>100$~kpc) showing where they lie in the Y--Z plane and on the energy-angular momentum diagram.
\label{fig:EAng}}
\end{figure*}

The large tail of dwarfs with highly eccentric orbits seen in model D is also reflected in the energy-angular momentum diagram (Figure \ref{fig:EAng}), with these dwarfs occupying high angular momentum, high energy positions on this diagram.
Relatively few unbound dwarfs occupy the same position on this diagram as dwarfs that end up bound, and any detected dwarfs that occupy this position will be less likely to have fallen in with the LMC on a previous orbit.

\section{Discussion}
The appearance of a disk-of-satellites around the Milky Way, has been associated with the infall of a group of dwarf galaxies \citep{Li2008} and in particular the infall of the Magellanic system \citep{D'Onghia2008}.
This scenario---an association of dwarfs falling onto the Milky Way---has been found to require a much more compact distribution than is seen in dwarf associations in the local group \citep{Metz2009}.
These bound associations were also unstable when orbiting a larger Galaxy.
Few dwarfs who would spend their entire undisturbed orbit within the tidal radius remain bound after two pericentre passages, and almost half are lost within $500$~Myr of the first pericentre passage.
Here we examine the properties of dwarfs that were bound to the LMC in the past, and whether this scenario could be responsible for some of the dwarf galaxies seen today.
As no dwarfs seen today are bound to the LMC, possibly excluding the SMC, any dwarfs associated in the past with the Magellanic system were likely to have been high in the potential well.
If there were multiple small dwarfs bound to the LMC, they were unlikely to have interacted, with no difference seen between models which had only one dwarf associated and those with four.
The presence of smaller dwarfs, which possessed insufficient gas to form dwarf galaxies will be more numerous and may have an effect on the orbit of these dwarfs.
This lack of interaction may arise from the LMC stabilizing the orbits of dwarfs against minor perturbations.
Interactions with larger galaxies, such as dwarfs interacting with LMC sized galaxies, or the LMC/SMC interactions can produce noticeable differences in the orbits and morphology of dwarf galaxies \citep{D'Onghia2009,Besla2010}.

Of the dwarfs listed by \citet{Kroupa2005} as possibly part of the Magellanic system when falling in, only Sagittarius, Ursa Minor, Sculptor, Carina and Draco are located within the area that our model dwarfs end up.
Sextans and Fornax, located slightly outside the range of dwarfs, may also have fallen in with the Magellanic system having become unbound before the previous two apogalacticons.
Using the velocities from \citet{Lux2010} we examine the direction of travel of the dwarfs in the Y--Z plane.
Ursa Minor and Sagittarius are travelling clockwise in the Y--Z plane consistent with them falling in with the Magellanic system, while the remaining dwarfs have errors large enough to be consistent with both travelling clockwise and counter-clockwise.
Of the remaining dwarfs, Carina is unlikely to have interacted with the LMC in the last $3$~Gyr \citep{Pasetto2011} and correspondingly did not fall in with the Magellanic system.
Future observations, with precise proper motions (e.g. from GAIA) may exclude more dwarfs using this simple rotation test, or via their positions on an energy-angular momentum diagram, with the dwarfs from our models ending up in only a few regions of the diagram, some of which are exclusively bound to the LMC (of the dwarfs seen today, only the SMC may be bound).

The assumption of a spherically smooth and static potential for the Milky Way halo is known to be unrealistic.
A triaxial potential can have noticeable impacts on dwarf orbits over time \citep{Lux2010}, but this potential makes modelling the history of observed dwarfs backwards much harder.
On the other hand it may allow more of the dwarfs observed today the possibility of being associated with the Magellanic system at an earlier time.

Models in which the Magellanic system is on its first orbit of the Galaxy show a much reduced loss.
This is partially due to the perturbation of the LMC and SMC's orbits by dwarfs over long times.
The possiblity of dwarf companions therefore needs to be taken into account when determing the orbit of the LMC and SMC.
The LMC, and to a lesser extent the SMC, were assumed to be tidally stripped before entering the Galactic halo.
Tidal stripping that happens as the LMC enters the halo will increase dynamical friction and subsequently result in a lower velocity today than those present in model F.
As most dwarfs are lost when the SMC binds to the LMC \citep[an event that is likely to have happened after the LMC enters the Galactic halo; see][]{Boylan-Kolchin2011} or at perigalaction (approximately today), dwarfs in this scenario may be better represented by model C.

The restriction that dwarfs initially spend their entire orbit within the tidal radius of the LMC (with respect to the Galaxy) clearly leaves out bound orbits (those with negative energy with respect to the LMC).
Dwarfs on these extra-tidal orbits are extremely likely to be tidally stripped and and could also be the source of some dwarf galaxies that reside in the disk-of-satellites. 

\section{Conclusion}
We have modelled the orbits of a group of dwarf galaxies bound to the LMC at a previous apogalacticon using a Monte Carlo approach, varying the orbits within observational errors.
Dwarf galaxies bound to the LMC are unlikely to interact with each other, and a significant fraction become unbound from the Magellanic system after only half an orbit.
These dwarfs would be located around the Magellanic system today and likely be noticeably associated with it.
Dwarfs that were bound one and a half orbits ago---that is at the LMC's second-last apogalacticon---are over six times more likely to become unbound than remain bound, with many dwarfs being located either in an extended structure of orbits, or in a tight disk around the Galaxy.
This disk encompasses the locations of a number of dwarfs observed today around the Milky Way, so a number of these dwarfs may originally have fallen in with the Magellanic system and been captured by the Galaxy.
The common rotation direction of the dwarfs in this ring provides a test to rule out any counter-rotating dwarfs as originally associated with the Magellanic system.
The extended disk-of-satellites cannot be explained by the dwarfs being bound to the LMC within the last two apogalacticons, and may have another origin.
In addition, the anomalous velocity and position of Leo~I is not explained by this mechanism, with no dwarfs in any of the simulations approaching the position, let alone the velocity, of Leo~I.

\acknowledgements

J.B.-H. is supported by a Federation Fellowship from the Australian Research Council.

\end{document}